\documentstyle[amssymb,aps,multicol,prl]{revtex}
\begin{document}
\title{Investigating The Vortex Melting Phenomenon In BSCCO Crystals Using Magneto-Optical Imaging Technique}
\author{A. Soibel$^{1}$, S. S. Banerjee$^{1}$,  Y. Myasoedov$^{1}$, M. L. Rappaport$^{1}$, E. Zeldov$^{1}$, S. Ooi$^{2}$, T. Tamegai$^{2,3}$}
\address{$^{1}$Department of Condensed Matter Physics, The Weizmann Institute of Science, Rehovot 76100,
Israel; $^{2}$Department of Applied Physics, The
University of Tokyo, Hongo, Bunkyo-ku, Tokyo 113-8656, Japan; $^{3}$ CREST, Japan Science and Technology Corporation
(JST), Japan} 
\maketitle

\begin{abstract}
Using a novel differential magneto-optical imaging technique we
investigate the phenomenon of vortex lattice melting in crystals of Bi$_2$Sr$_2$CaCu$_2$O$_8$
(BSCCO). The images of melting reveal complex patterns in the formation and evolution of the vortex 
solid-liquid interface with varying field ($H$) / temperature ($T$). We believe that the complex melting patterns are due to a random distribution of material disorder/inhomogeneities across the sample, which create fluctuations in the local melting temperature or field value. To study the fluctuations in the local melting temperature / field, we have constructed maps of the melting landscape $T_m(H,r)$, viz., the melting temperature ($T_m$) at a given location ($r$) in the sample at a given field ($H$). A study of these melting landscapes reveals an unexpected feature: the melting landscape is not fixed, but changes rather dramatically with varying field and temperature along the melting line. It is concluded that the changes in both the scale and shape of the landscape result from the competing contributions of different
types of quenched disorder which have opposite effects on the local melting transition.
\end{abstract}

\begin{center}

\end{center}

\vskip0.3 truecm
\noindent
{\bf Introduction :}
General arguments \cite{Imry} suggest that weak disorder results in a rounding of a first-order transition while extensive disorder transforms it into second order, yet the details of this ubiquitous process on the atomic level are still not clear. The vortex lattice in superconductors provides a unique and a relatively easy way to study this first-order transition \cite{Blatter,Zeldov1} over a wide range of particle densities, by just varying the magnetic field at different temperatures. By use of a novel differential magneto-optical imaging technique a direct experimental visualization of the melting process in a disordered system is obtained, revealing complex melting patterns. From  the images of melting at different fields and temperatures we investigate the behavior of the local melting temperature at different fields, viz., we construct the melting landscape $T_m(H,r)$. Such a map has revealed a competition between different types of pinning arising out of sample inhomogeneities / disorder. The different types of pinning have opposite effect on the local value of melting $T/H$.
\vskip 0.15truecm
\noindent
{\bf The Differential Magneto-Optical Technique :}
Briefly, the MO imaging is achieved by placing a garnet indicator film on a sample. Linearly polarized light undergoes Faraday rotation in the indicator and reflected back through a crossed polarizer, resulting in a real-time image of the magnetic field distribution. 

Under equilibrium magnetization conditions in platelet-shaped samples in perpendicular applied field $H_a\parallel$z, the dome-shaped profile of the internal field \cite{Zeldov2}, which has a maximum in the center, results in the central part of the sample reaching the melting field $H_m(T)$ first upon increasing field or temperature. We should therefore expect the nucleation of a small round `puddle' of vortex liquid in the center of the sample, surrounded by vortex solid. Due to the first-order nature of the transition, the vortex-lattice melting is associated with a discontinuous step in the equilibrium magnetization\cite{Zeldov1}, $4\pi \Delta M = \Delta (B-H)$. Since in our geometry the field $H$ is continuous across the solid-liquid interface, the induction B in the liquid is thus enhanced by $\Delta B$ relative to the solid. In BSCCO crystals $\Delta B$ is typically 0.1 to 0.4 G \cite{Zeldov1}. Conventional magneto-optical (MO) imaging techniques \cite{Indenbom,Duran,Welp} (see Fig. 1 caption) cannot resolve such small field differences. We have therefore devised the following differential method. 

A MO image is acquired by averaging typically ten consecutive CCD images at fixed $H_a$ and $T$. Then $H_a$ is increased by $\delta H_a << H_a$, or $T$ is increased by $\delta T << T$, and a second averaged image is obtained. The difference between the two images is stored as the differential image. This process is repeated and averaged typically 100 times, yielding field resolution of about 30 mG, which is typically two orders of magnitude better than the standard MO method. By recording the differential images in a sequence of fields or temperatures a `movie' of the melting process is obtained \cite{Soibel}. 

In the absence of disorder, as the field is increased by $\delta H_a$, the radius of the vortex-liquid puddle in the center of the sample should increase by $\delta R$, determined by the gradient of the dome-shaped profile $H(x,y)$. The differential image in this case should show a bright ring on a dark background, which indicates the location of the expanding solid-liquid interface. In the rest of the image almost no change in the field should occur, except the uniform background signal of $\delta H_a$. The intensity of the ring is $\Delta B$ above the background\cite{Morozov}, independent of $\delta H_a$, whereas the width of the ring reflects the distance $\delta R$ over which the interface expands due to field modulation $\delta H_a$. 
\vskip 0.15truecm
\noindent
{\bf Observation of Melting Patterns in BSCCO Crystals :}
Figure 1 presents several differential MO images of the vortex-lattice melting at $T$ = 60 K in one of the BSCCO crystals, which is initially in the vortex solid phase. At 159.5 Oe a small liquid puddle is nucleated, seen as a bright spot in the upper-right part. Note that in contrast to expectations, the puddle is not in the center of the sample, nor is it round, but instead a rather rough shape of the vortex liquid domain is observed. As the field is further increased a ring-like bright object is obtained, which is the solid-liquid interface separating the liquid from the surrounding vortex solid. Both the shape and the width of the ring are highly nonuniform. At 165 Oe a `tongue' of the liquid protrudes sharply to the left side. By 168 Oe the upper part of the sample is entirely in the liquid phase, with a rough interface separating the liquid from the narrow solid region at the bottom. We shall attempt to understand the influence of material disorder/inhomogeneities across the sample on local melting properties.
\vskip 0.15truecm
\noindent
{\bf Investigating the Melting Landscape in BSCCO :}
Since the differential MO technique enables us to spatially resolve the location of the vortex solid-liquid interface, we thus attempt to deduce from the melting images the spatial ({\it r}) variations in the melting temperature ($T_m$) at different fields ({\it H}) and at different locations ($r$) in the sample, i.e., $T_m(H,r)$. There have been numerous studies in the past relating to the effect of disorder on the melting phenomenon. Weak point disorder is expected to shift the melting transition to lower temperature, while preserving the first order nature \cite{Khaykovich,Paulius}, while correlated disorder shifts the melting transition to higher temperatures \cite{Khaykovich2,Kwok2}. Oxygen doping is another parameter that, apart from affecting the $T_c$ of the sample, changes the material anisotropy, which in turn significantly changes the slope of the melting line \cite{Khaykovich3}. All these different parameters not only determine the location of the mean field melting transition line $T_m(H)$, but should also lead to significant fluctuations in the value of $T_m(H,r)$.

In Figs. 2a and 2b we present a spatial distribution of the
liquid regions nucleating in the sample for different $T$ at a
fixed $H_a$, the various colors indicating areas which melt during a 
0.25 K increment in $T$. Material disorder modifies the local
melting temperature, thus forming a complicated $T_m(H,
{\it r})$ landscape. Figures 2a and 2b can thus be viewed as
`topographical maps' of the melting landscape at $H_a$ of 20 and
75 Oe, respectively. The minima points or the valleys of the
landscape (blue) melt first whereas the peaks of the landscape (red)
melt last. The two landscapes in Fig. 2 are substantially
different. In addition to the significant change in the
characteristic length scale and roughness of the landscape, there
are many regions in the sample that show qualitatively different
properties. By comparing the figures for 20 Oe and 75 Oe, one can 
clearly see how peaks of the melting landscape change into valleys.
For example, at 20 Oe the valley in the form of an arc
along the `O-O' dashed line has three long and narrow blue
segments, while at 75 Oe, the blue minima have the form of rather
circular spots. Also, at 75 Oe to the right of the `O-O' valley, a number of extended peaks colored yellow and orange can be seen, while at the same locations at 20 Oe we find blue and green valleys. Also, importantly, the width
of the transition or the valley-to-peak height, changes
significantly. At 20 Oe the entire sample melts within about 1 K,
whereas at 75 Oe the melting process spans almost twice this
range. We have performed a more quantitative analysis by investigating the melting behavior at several points, eg., at points A and B in Fig. 2 \cite{Soibel2}. We have found that above 85 K the point B systematically melts about 0.5 K below point A and below 85 K, the point B melts up to 2 K above point A. Numerous other points also show similar crossing behavior in their local melting temperatures. This implies that the valleys in the melting landscape at low fields may turn into peaks at higher fields. 
We try to understand the crossing of melting landscape using the generic expression of the melting
transition \cite{BlatIvl}, $H_m(T,\textbf{r})=H_0(\textbf{r})(1-T/T_c(\textbf{r}))^{\alpha
(\textbf{r})}$. We now set $\alpha (\textbf{r})= \alpha $ for simplicity 
and we introduce spatial variations in $T_c$ and $H_0$ with $T_c(\textbf{r})= T_c+ \Delta T_c(\textbf{r})$ and $H_0(\textbf{r}) = H_0+\Delta H_0(\textbf{r})$, we rewrite $H_m(T, \textbf{r})=H_m(T)+\Delta
H_m(T, \textbf{r})$. Using these we get the expression $\Delta H_m(T, \textbf{r})\simeq [\Delta H_0(\textbf{r})+\alpha H_0 \Delta T_c(\textbf{r})/ (T_c~-~T)](1-T/T_c)^{\alpha}$ describing the disorder-induced melting landscape. Out of the two terms in the expression, we see that $\Delta H_o(r)$ dominates at low T, while the second term containing $\Delta T_c(r)$ dominates close to $T_c$.
It is the competition arising out of the two terms which causes the crossing over in the melting landscape. To illustrate our point further consider the following : At high temperatures $\Delta T_c$ variations should 
modify the local critical field $H_{c1}(\textbf{r})$. Figure 2c
shows a high-sensitivity image of the initial field penetration
into the sample at $H_a$=2 Oe and $T$=89 K. A strong correlation
between the field penetration form and the melting patterns in Fig. 2a is readily visible. For a more accurate comparison Fig. 2d
presents a superposition of the penetration image with the melting
patterns at 20 Oe. The color in the image is given by the melting
contours, while the brightness is defined by the penetration
field. It is clearly seen that most of the macroscopic blue
regions of  liquid nucleation coincide with the bright areas where
the field penetrates first. In particular, the correspondence
between the arc structures of the penetration field and the
melting contours is striking. This correspondence indicates that
the melting propagation at low fields is indeed governed by the
local variations in $T_c$. A comparison of the penetration image and the melting contours
at $H_a$ = 75 Oe reveals surprisingly large anti-correlation
behavior: The regions into which the field penetrates first are often
the last ones to melt, for
example, the `P-P' strip is bright in Fig. 2c but is
mainly yellow and orange in Fig. 2b. The anti-correlation behavior
causes the observed crossing of the local melting lines as discussed above.
\vskip 0.15truecm
\noindent
{\bf Summary :}
We have studied the vortex melting phenomenon in BSCCO using the sensitive differential magneto-optical imaging technique which enables the observation of local jumps in magnetization of the order of 100 to 300 mG. Contour maps of the nucleation and the evolution of the vortex solid-liquid interface as a function of field or temperature have been constructed. We surmise that competing contributions of different types of quenched random disorder which locally affect $T_c$ or $H_0$ produce opposite effects on the local melting transition, and we recognize them to be a crucial ingredient in determining the complexity of the melting patterns. 
\vskip 0.05truecm
\noindent
{\bf Acknowledgements:}
This work was supported by the Israel Science Foundation and Center of Excellence Program, by Minerva Foundation, Germany, by the Ministry of Science, Israel, and by the Grant-in-Aid for Scientific Research from the Ministry of Education, Science, Sports and Culture, Japan. E. Z. acknowledges the support by the Fundacion Antorchas - Weizmann Institute of Science Collaboration Program.

\vskip 0.5truecm
{\bf Figure Captions}
\vskip 0.8truecm
\noindent
{\bf Fig.1.} Vortex lattice melting process in a small BSCCO crystal. Differential MO images of the melting process in a BSCCO crystal ($T_c=90$ K) of area 0.35 $\times$ 0.27 mm$^2$ at $T$=60 K and $H_a \parallel $ c-axis. The gray scale from black to white spans a field range of 0.2 G. The region outside the sample is bright. The differential images are obtained by subtracting the image at field 
$Ha$ from the image at $H_a + \delta H_a$, with $\delta H_a=1$ Oe. A detailed discussion of the images can be found in the text.
\vskip 1.5truecm
\noindent
{\bf Fig.2.} The contours of the melting propagation in BSCCO crystal ($T_c = 91$K, dimensions 1 $\times$ 1 $\times$ 0.05 mm$^3$)
at $H_a$ = 20 Oe (a)  and 75 Oe (b). The color code indicates the expansion of the liquid domains as the temperature is increased in 0.25 K steps. The onset of melting at $T_m^{on}$ = 86.25 K in (a) and at 74.25 K in (b). (c) Differential magneto-optical image of the magnetic field penetration at $H_a$ = 2 Oe, $T$ = 89 K, $\delta H_a$ = 1 Oe. (d) Superposition of (a) and (c).
\end{document}